\newif\ifLNCS
\LNCSfalse
\newif\ifHAL 
\HALtrue
\ifLNCS \documentclass{llncs}\usepackage{minted}
\else \ifHAL
\documentclass{article}
\usepackage{color}
\usepackage{alltt}
\newenvironment{minted}[1]%
{\begin{samepage}\begin{alltt}}%
{\end{alltt}\end{samepage}}
\else 
\documentclass[a4paper]{amsart}\usepackage{minted}\usepackage{color}
\fi
\fi

\usepackage{fullpage}

\usepackage[utf8]{inputenc}
\usepackage[T1]{fontenc}
\usepackage[english]{babel}

\usepackage[all]{xy}
\usepackage{qsymbols,xparse}
\usepackage{amsmath,amstext}
\usepackage{graphicx}

\usepackage{url}
\usepackage{breakurl} 
\usepackage{nicefrac}

\ifLNCS \else 
\usepackage{amsthm}
\theoremstyle{definition}

\fi

\newcommand{\A}{\mathcal{A}}

\renewcommand{\L}{\mathcal{L}}

\newcommand{\ANF}{\mathcal{A}^{nf}}
\newcommand{\ANE}{\mathcal{A}^{ne}}
\newcommand{\LNF}{\mathcal{L}^{nf,`n}}
\newcommand{\LNEZ}{\mathcal{L}^{ne,0}}
\newcommand{\LNFZ}{\mathcal{L}^{nf,0}}
\newcommand{\LNE}{\mathcal{L}^{ne,`n}}

\newcommand{\m}{\mathbf{m}}
\newcommand{\uu}{\mathbf{u}}
\newcommand{\inc}[2]{#2^{\uparrow #1}}

\newcommand{\nat}{\mathbb{N}}

\newcommand{\tail}{\mathsf{tail}\,}

\DeclareDocumentCommand{\gL}{O{\infty}}{L_{#1}(z)}
\DeclareDocumentCommand{\cgL}{O{\infty}}{\mathcal{L}_{#1}}

\DeclareDocumentCommand{\agL}{O{\infty}}{M_{#1}(z)}
\DeclareDocumentCommand{\acgL}{O{\infty}}{\mathcal{M}_{#1}}

\newcommand{\lf}[2]{{\raisebox{.8pc}{\tiny
\xymatrix@=.05pc{&{#2}\ar@{-}[dl]\\{#1}}}}}

\newcommand{\ri}[2]{{\raisebox{.8pc}{\tiny
\xymatrix@=.05pc{{#1}\ar@{-}[dr]\\&{#2}}}}}

\definecolor{vertfonce}{rgb}{0,.5,0}
\definecolor{mauve}{rgb}{1,0,1}
\definecolor{rougefonce}{cmyk}{.3,1,.3,0}
\definecolor{cyanp}{cmyk}{.5,.3,0,0}
\definecolor{yellow}{cmyk}{0,0,.7,0}
\definecolor{beige}{cmyk}{0,.2,.7,0}
\definecolor{brun}{cmyk}{0,.5,.7,0}
\definecolor{brunfonce}{cmyk}{.3,.75,.75,.15}
\newcommand{\rouge}[1]{{\color{red} #1}}
\newcommand{\pourpre}[1]{{\color{rougefonce} #1}}

\newcommand{\bl}[1]{\textcolor{blue}{#1}}
\newcommand{\verdir}[1]{{\color{vertfonce} #1}}

  {\color{blue}}
  {}
  {\color{brunfonce}}
  {}

  {\color{red}}
  {}

  \newcommand{\Abstract}{%
    Affine $`l$-terms are $`l$-terms in which each bound variable occurs
    at most once and linear $`l$-terms are $`l$-terms in which each bound
    variable occurs once. and only once.  In this paper we count the
    number of closed affine $`l$-terms of size $n$, closed linear
    $`l$-terms of size $n$, affine $`b$-normal forms of size $n$ and
    linear $`b$-normal forms of size $n$, for different ways of
    measuring the size of $`l$-terms.  From these formulas, we show how we
    can derive programs for generating all the terms of size $n$ for each
    class.  The foundation of all of this is specific data structures,
    which are contexts in which one counts all the holes at each level of
    abstractions by $`l$'s.}

\ifLNCS
\title{Quantitative aspects of \\linear and affine closed lambda terms}
\author{Pierre Lescanne}
\institute{University of Lyon\\
        \'Ecole normale sup\'erieure de Lyon\\
        LIP (UMR 5668 CNRS ENS Lyon UCBL INRIA)\\
        46 all\'ee d'Italie, 69364 Lyon, France\\[12pt]
        \textsf{pierre.lescanne@ens-lyon.fr}}
\date{\today}
\begin{document}
\maketitle
\begin{abstract}
  \Abstract

\end{abstract}
\else
\ifHAL
\title{Quantitative aspects of \\linear and affine closed lambda terms}
\author{Pierre Lescanne\\
  University of Lyon\\
        \'Ecole normale sup\'erieure de Lyon\\
        LIP (UMR 5668 CNRS ENS Lyon UCBL INRIA)\\
        46 all\'ee d'Italie, 69364 Lyon, France\\[12pt]
        \textsf{pierre.lescanne@ens-lyon.fr}}
\date{\today}
\begin{document}
\maketitle
\begin{abstract}
  \Abstract

\end{abstract}
\else
\begin{document}
\title{Quantitative aspects of \\linear and affine closed lambda terms}
\author{Pierre Lescanne}
\address{University of Lyon\\
  \'Ecole normale sup\'erieure de Lyon\\
  LIP (UMR 5668 CNRS ENS Lyon UCBL INRIA)\\
  46 all\'ee d'Italie, 69364 Lyon, France}
\email{pierre.lescanne@ens-lyon.fr} \date{\today}
\begin{abstract}
  \Abstract

\end{abstract}
\maketitle
\fi \fi
\section{Introduction}
\label{sec:intro}

The $`l$-calculus~\cite{Barendregt84} is a well known formal system
designed by Alonzo Church~\cite{Church40} for studying the concept of
function. It has three kinds of basic operations: variables, application
and abstraction (with an operator $`l$ which is a binder of variables).
We assume the reader familiar with the $`l$-calculus and with de Bruijn
indices.\footnote{If the reader is not familiar with the $`l$-calculus, we
  advise him to read the introduction
  of~\cite{DBLP:journals/jfp/GrygielL15}, for instance.}

In this paper we are interested in terms in which bound variables occur
once.  A \emph{closed $`l$-term} is a $`l$-term in which there is no free
variable, i.e., only free variables.  An \emph{affine $`l$-term} (or BCK
term) is a $`l-$term in which bound variables occur at most once.  A
\emph{linear} $`l$-term (or BCI term) is a $`l$-term in which bound
variables occur once and only once.

In this paper we propose a method for counting and generating (including
random generation) linear and affine closed $`l$-terms based on a data
structure which we call \emph{SwissCheese} because of its holes.  Actually
we count those $`l$-terms up-to $`a$-conversion.  Therefore it is adequate
to use de Bruijn indices~\cite{NGDeBruijn108}, because a term with de Bruijn
indices represents an $`a$-equivalence class.  An interesting aspect of
these terms is the fact that they are simply
typed~\cite{hindley97:_basic_simpl_theor,DBLP:journals/tcs/Hindley89}.
For instance, generated by the program of Section~\ref{sec:genF}~\ref{sec:comp}, there
are $16$ linear terms of natural size $8$:
\begin{footnotesize}
  \begin{displaymath}
    (`l0~(`l0~`l0)) \quad (`l0~`l(`l0~0)) \quad (`l0~`l(0~`l0)) \quad ((`l0~`l0)~`l0) \quad (`l(`l0~0)~`l0) \quad (`l(0~`l0)~`l0) \quad `l(`l0~(`l0~0)) \quad `l(`l0~(0~`l0))
  \end{displaymath}
  \begin{displaymath}
    `l((`l0~`l0)~0) \quad `l(`l(`l0~0)~0) \quad `l(`l(0~`l0)~0) \quad `l(0~(`l0~`l0)) \quad `l(0~`l(`l0~0)) \quad `l(0~`l(0~`l0)) \quad `l((`l0~0)~`l0) \quad `l((0~`l0)~`l0)
  \end{displaymath}
\end{footnotesize}
written with explicit variables
\begin{footnotesize}
\newcommand{\lO}{`l x . x}
  \begin{displaymath}
        \lO~(\lO~\lO) \quad \lO~`l y . ( \lO~y) \quad \lO~`l y . ( y~\lO)  \quad  (\lO~\lO)~\lO 
        \end{displaymath}
        \begin{displaymath}
        `l y . ( \lO~y)~\lO \quad `l y . ( y~\lO)~\lO \quad
          `l y . ( \lO~(\lO~y)) \quad `l y . ( \lO~(y~\lO)) 
        \end{displaymath}
  \begin{displaymath}
    `l y . ( (\lO~\lO)~y) \quad `l y . ( `l z . ( \lO~z)~y) \quad `l y . ( `l z . ( z~\lO)~y) \quad
    `l y . ( y~(\lO~\lO)) 
    \end{displaymath}
    \begin{displaymath}      
`l y . ( y~`l z . ( \lO~z)) \quad `l y . ( y~`l z. ( z~\lO)) \quad `l y . ( (\lO~y)~\lO) \quad `l y . ( (y~\lO)~\lO)
  \end{displaymath}
\end{footnotesize}
and there are $25$ affine terms of natural size $7$:
\begin{footnotesize}
  \begin{displaymath}
    (`l0~`l`l1) \quad (`l0~`l`l`l0) \quad (`l`l0~`l`l0) \quad (`l`l1~`l0) \quad (`l`l`l0~`l0) \quad `l(`l`l1~0) \quad `l(`l`l`l0~0) \quad `l(0~`l`l1) 
  \end{displaymath}
  \begin{displaymath}
    ~`l(0~`l`l`l0) \quad `l(`l0~`l1) \quad `l(`l1~`l0) \quad `l`l(`l0~1) \quad `l`l(1~`l0) \quad `l(`l0~`l`l0) \quad `l(`l`l0~`l0) \quad `l`l(`l`l0~0)
  \end{displaymath}
  \begin{displaymath}
    ~`l`l(0~`l`l0) \quad `l`l`l(0~1) \quad `l`l`l(1~0) \quad `l`l`l`l2 \quad `l`l(`l0~`l0) \quad `l`l`l(`l0~0) \quad `l`l`l(0~`l0) \quad `l`l`l`l`l1 \quad `l`l`l`l`l`l0
  \end{displaymath}
\end{footnotesize}

The Haskell programs of this development are on GitHub: \url{https://github.com/PierreLescanne/CountingGeneratingAfffineLinearClosedLambdaterms}.
\subsection*{Notations}
\label{sec:notation}
In this paper we use specific notations.

Given a predicate $p$, the Iverson notation written $[p(x)]$ is the
function taking natural values which is~$1$ if $p(x)$ is true and which is
$0$ if $p(x)$ is false.

Let $\m`:\nat^{p}$ be the $p$-tuple $(m_0,...,m_{p-1})$. In
Section~\ref{sec:genF}, we consider also infinite tuples. Thus
$\m`:\nat^{`w}$ is the sequence $(m_0, m_1, ...)$.  Notice in the case of
infinite tuples, we are only interested in infinite tuples equal to $0$
after some index.
\begin{itemize}
\item $p$ is the \emph{length} of $\m$, which we write also
  $\mathsf{length}~\m$
\item The $p$-tuple $(0,..., 0)$ is written $0^p$. $0^{`w}$ is the
  infinite tuple made of $0$'s.
\item The \emph{increment} of a $p$-tuple at $i$ is:
  \begin{displaymath}
    \inc{i}{\m} = \mathbf{n}`:\nat^{p} \mathrm{~where~} n_j = m_j
    \mathrm{~if~} j\neq i \mathrm{~and~} n_i = m_i + 1
  \end{displaymath}
\item Putting an element $x$ as \emph{head} of a tuple is written
  \begin{displaymath}
    x:\m = x : (m_0, ...) = (x,m_0, ...)
  \end{displaymath}
  \textsf{tail} removes the head of a tuple:
  \begin{displaymath}
    \mathsf{tail}(x:\m) = \m.
  \end{displaymath}

\item $`(+)$ is the componentwise addition on tuples.
\end{itemize}

\section{SwissCheese}
\label{sec:swc}

The basic concept is this of \textbf{$\m$-SwissCheese} or
\textbf{Swisscheese of characteristic $\m$} or simply \textbf{SwissCheese}
if there is no ambiguity on $\m$.  A $\m$-SwissCheese or a SwissCheese of
characteristic $\m$, where $\m$ is of length~$p$) is, a $`l$-term with
holes at $p$ levels, which are all counted, using $\m$.  The $p$ levels of
holes are $\Box_0$,... $\Box_{p-1}$. A hole $\Box_i$ is meant to be a
location for a variable at level $i$, that is under $i$ $`l$'s.  According
to the way bound variables are inserted when creating abstractions (see
below), we consider linear or affine SwissCheeses.  The holes have size
$0$. An $\m$-SwissCheese or a SwissCheese of characteristic $\m$ has $m_0$
holes at level~$0$, $m_1$ holes at level $1$, ...  $m_{p-1}$ holes at
level $p-1$.  Let $l_{n,\m}$ (resp. $a_{n,\m}$) count the linear
(resp. the affine) $\m$-SwissCheese of size $n$.  $l_{n,\m} = l_{n,\m'}$
and $a_{n,\m} = a_{n,\m'}$ if $\m$ is finite, $\mathsf{length}~\m \ge n$,
$m_i= m'_i$ for $i\le \mathsf{length}~\m$, and $m'_i=0$ for $i >
\mathsf{length}~\m$.  $l_{n,0^n}$ (resp. $a_{n,0^n}$) counts the closed
linear (resp. the closed affine) $`l$-terms.

\subsection{Growing a SwissCheese}
\label{sec:grow}

\begin{figure}[t!]
  \centering
  \begin{displaymath}
    c_1 = \xymatrix@R=3pt@C=3pt{&`l\ar@{-}[d]\\&{@}\ar@{-}[dl]\ar@{-}[dr]\\{\Box_1}&&0}%
    \qquad
    c_2 = \xymatrix@R=3pt@C=3pt{&@\ar@{-}[dl]\ar@{-}[dr]\\`l\ar@{-}[d]&&\Box_0\\\Box_1}
    \qquad\qquad
    \xymatrix@R=3pt@C=3pt{&@\ar@{-}[dl]\ar@{-}[dr]\\c_1&&c_2} = 
    \xymatrix@R=3pt@C=3pt{&&&@\ar@{-}[dll]\ar@{-}[drr]\\ %
      &`l\ar@{-}[d]&&&&@\ar@{-}[dl]\ar@{-}[dr]&\\ %
      &@\ar@{-}[dl]\ar@{-}[dr]&&&`l\ar@{-}[d]&&\Box_0\\ %
      \Box_1&&0&&\Box_1}
  \end{displaymath}
  \caption{Building a SwissCheese by application}
  \label{fig:app}
\end{figure}
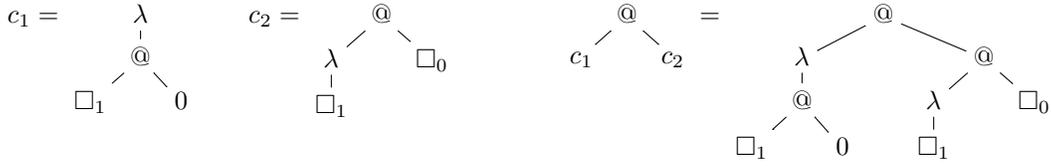

Given two SwissCheeses, we can build a SwissCheese \emph{by application} like in
Fig~\ref{fig:app}.
In Fig.~\ref{fig:app}, $c_1$ is a $(0,1,0,0,0)$-SwissCheese, $c_2$ is a
$(1,1,0,0,0)$-SwissCheese and $c_1 @ c_2$ is a $(1,2,0,0,0)$-SwissCheese.
Said otherwise, $c_1$ has characteristic $(0,1,0,0,0)$, $c_2$ has
characteristic $(1,1,0,0,0)$ and $c_1 @ c_2$ has characteristic
$(1,2,0,0,0)$.  According to what we said, $c_1 @ c_2$ has characteristic
$(1,2)$ as well as characteristic $(1,2,0,0,...)$ (a tuple starting with
$1$, followed by $2$, followed by infinitely many $0$'s).  We could also
say that $c_1$ has characteristic $(0,1)$ and $c_2$ has characteristic
$(1,1)$ making $@$ a binary operation on SwissCheeses of length $2$
whereas previously we have made $@$ a binary operation on SwissCheeses of
length $5$.  In other words, when counting SwissCheeses of characteristic
$\m$, the trailing $0$'s are irrelevant.  In actual computations, we make
the lengths of characteristics consistent by adding trailing $0$'s to
too short ones.

\begin{figure}[t!]
  \centering
  \begin{displaymath}
    \xymatrix@R=3pt@C=3pt{\bl{`l}\ar@{-}[d]\\c_2} = %
    \xymatrix@R=3pt@C=3pt{&\bl{`l}\ar@{-}[d]\\&@\ar@{-}[dl]\ar@{-}[dr]\\`l\ar@{-}[d]&&\Box_1\\\Box_2}%
    \qquad\qquad
    \xymatrix@R=3pt@C=3pt{&&&\verdir{`l}\ar@{-}[d]\\&&&@\ar@{-}[dll]\ar@{-}[drr]\\ %
      &`l\ar@{-}[d]&&&&@\ar@{-}[dl]\ar@{-}[dr]&\\ %
      &@\ar@{-}[dl]\ar@{-}[dr]&&&`l\ar@{-}[d]&&\Box_1\\ %
      \Box_2&&0&&\rouge{\Box_1}} = %
    \xymatrix@R=3pt@C=3pt{&&&\verdir{`l}\ar@{-}[d]\\&&&@\ar@{-}[dll]\ar@{-}[drr]\\ %
      &`l\ar@{-}[d]&&&&@\ar@{-}[dl]\ar@{-}[dr]&\\ %
      &@\ar@{-}[dl]\ar@{-}[dr]&&&`l\ar@{-}[d]&&\Box_1\\ %
      \Box_2&&0&&\rouge{S}\ar@[red]@{-}[d]\\
      &&&& \rouge{0}}
  \end{displaymath}
  \caption{Abstracting SwissCheeses without and with binding}
  \label{fig:abs}
\end{figure}
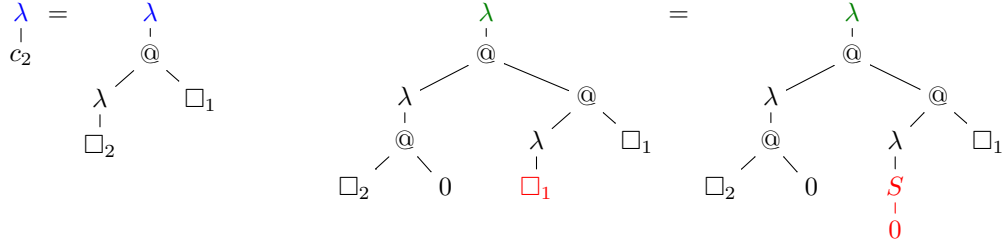

Given a SwissCheese, there are two ways to grow a SwissCheese to make
another SwissCheese \emph{by abstraction}.
\begin{enumerate}
\item We put a $`l$ on the top of a $\m$-SwissCheese $c$. This increases
  the levels of the holes: a hole $\Box_i$ becomes a hole $\Box_{i+1}$.
  $`l c$ is a $(0:\m)$-SwissCheese. See Fig~\ref{fig:abs} on the left.
  This way, no index is bound by the top $`l$, therefore this does not
  preserve linearity (it preserves affinity however). Therefore this
  construction is only for building affine SwissCheeses, not for building
  linear SwissCheeses.  In Figure~\ref{fig:abs} (left), we colour the
  added $`l$ in blue and we call it \emph{abstraction with \underline{no}
    binding}.
\item In the second method for growing a SwissCheese by abstraction, we
  select first a hole $\Box_i$, we top the SwissCheese by a $`l$, we
  increment the levels of the other holes and we replace the chosen hole by
  $S^i 0$.  In Figure~\ref{fig:abs} (right), we colour the added $`l$ in
  green and we call it \emph{abstraction \underline{with} binding}.
\end{enumerate}

\subsection{Measuring SwissCheese}
\label{sec:mes}

We considers several ways of measuring the size of a SwissCheese derived
from what is done on $`l$-terms.  In all these sizes, applications $@$ and
abstractions $`l$ have size $1$ and holes have size $0$. The differences
are in the way variables are measured.
\begin{itemize}
\item Variables have size $0$, we call this \textbf{variable size $0$}.
\item Variables have size $1$, we call this \textbf{variable size $1$ }.
\item Variables (or de Bruijn indices) $S^i 0$ have size $i+1$, we call
  this \textbf{natural size}.
\end{itemize}

\section{Counting linear closed terms}
\label{sec:affine}

We start with counting linear terms since they are slightly simpler.  We
will give recursive formulas first for the numbers $l^{`n}_{n,\m}$ of
linear SwissCheeses of natural size $n$ with holes set by $\m$, then for
the numbers $l^0_{n,\m}$ of linear SwissCheeses of size $n$, for variable
size $0$, with holes set by $\m$, eventually for the numbers $l^1_{n,\m}$
of linear SwissCheeses of size $n$, for variable size $1$, with holes set
by $\m$.  When we do not want to specify a chosen size, we write only
$l_{n,\m}$ without superscript.

\subsection{Natural size}
\label{sec:natLin}

First let us count linear SwissCheeses with natural size. This is given by
the coefficient $l^{`n}$ which has two arguments: the size $n$ of the
SwissCheese and a tuple $\m$ which specifies the number of holes of each
level.  In other words we are interested in the quantity $l_{n,\m}^{`n}$.  We
assume that the length of $\m$ is $p$, greater than $n$.
\begin{description}
\item[$\mathbf{n = 0}$] whatever size is considered, there is only one
  SwissCheese of size $0$ namely $\Box_0$. This means that the number of
  SwissCheeses of size $0$ is $1$ if and only if $\m = (1, 0, 0 ,...)$:
  \begin{displaymath}
    l_{0,\m}^{`n} = l_{0,\m}^0 = l_{0,\m}^1 = [m_0 = 1 \wedge \bigwedge_{j=1}^{p-1} m_j = 0]
  \end{displaymath}
\item[$\mathbf{n \neq 0}$ and application] if a $`l$-term of size $n$ has
  holes set by $\m$ and is an application, then it is obtained from a $`l$
  term of size $k$ with holes set by $\mathbf{q}$ and a $`l$ term of size
  $n-k-1$ with holes set by $\mathbf{r}$, with $\m=\mathbf{q}`(+)
  \mathbf{r}$:
  \begin{displaymath}
    \sum_{\mathbf{q} `(+) \mathbf{r}\,=\mathbf{m}}\ %
    \sum_{k=0}^n\, l_{k,\mathbf{q}}\,  l_{n-1-k,\mathbf{r}}
  \end{displaymath}
\item[$\mathbf{n\neq 0}$ and abstraction with binding] consider a level
  $i$, that is a level of hole $\Box_i$.  In this hole we put a term
  $S^{i-1}\,0$ of size $i$.  There are $m_i$ ways to choose a hole
  $\Box_i$. Therefore there are
  \begin{math}
    m_i\, l^{`n}_{n-i-1,\mathbf{m}}
  \end{math}
  SwissCheeses which are abstractions with binding in which a $\Box_i$ has
  been replaced by the de Bruijn index $S^{i-1}\,0$ among
  $l^{`n}_{n,0:\mathbf{m^{\downarrow i}}}$ SwissCheeses, where
  $\mathbf{m^{\downarrow i}}$ is $\mathbf{m}$ in which $m_i$ is
  decremented.  We notice that this refers only to an $\m$ starting with
  $0$. Hence by summing over $i$ and adjusting $\m$, this part contributes
  as:
  \begin{displaymath}
    \sum_{i=0}^{p-1} (m_i + 1)\ l^{`n}_{n-i,\inc{i}{\m}}
  \end{displaymath}
  to $l^{`n}_{n+1,0:\m}$.
\end{description}
We have the following recursive definitions of $l^{`n}_{n,\m}$:
\begin{eqnarray*}
  l^{`n}_{n+1,0:\mathbf{m}} &=&\sum_{\mathbf{q} `(+)
    \mathbf{r}\,=\,0:\mathbf{m}}\
  \sum_{k=0}^n\, l^{`n}_{k,\mathbf{q}}\,
  l^{`n}_{n-k,\mathbf{r}} + 
  \sum_{i=0}^{p-1} (m_i + 1)\ l^{`n}_{n-i,\inc{i}{\mathbf{m}}}\\
  l^{`n}_{n+1,(h+1):\mathbf{m}} &=& \sum_{\mathbf{q} `(+)
    \mathbf{r}=(h+1):\mathbf{m}}\
  \sum_{k=0}^{n}\  l^{`n}_{k,\mathbf{q}}\,  l^{`n}_{n-k,\mathbf{r}}
\end{eqnarray*}
Numbers of closed linear terms with natural size are given in Figure~\ref{fig:lin_until_10}.
\subsection{Variable size $0$}
\label{sec:lin0}

The only difference is that the inserted de Bruijn index has size $0$.
Therefore we have $m_i\, l^0_{n-1,\mathbf{m}}$ where we had
$m_i\, l^{`n}_{n-i-1,\mathbf{m}}$ for natural size.  Hence the formulas:
\begin{eqnarray*}
  l^0_{n+1,0:\mathbf{m}} &=&\sum_{\mathbf{q} `(+)
    \mathbf{r}\,=\,0:\mathbf{m}}\
  \sum_{k=0}^n\, l^0_{k,\mathbf{q}}\,  l^0_{n-k,\mathbf{r}} + 
  \sum_{i=0}^{p-1} (m_i + 1)\ l^0_{n,\inc{i}{\mathbf{m}}}\\
  l^0_{n+1,(h+1):\mathbf{m}} &=& \sum_{\mathbf{q} `(+)
    \mathbf{r}=(h+1):\mathbf{m}}\
  \sum_{k=0}^{n}\  l^0_{k,\mathbf{q}}\,  l^0_{n-k,\mathbf{r}}
\end{eqnarray*}
The sequence $l^0_{n,0^n}$ of the numbers of closed linear terms is
$0,1,0,5,0,60,0,1105,0,27120,0,828250, $ which is sequence
\textsf{A062980} in the \emph{On-line Encyclopedia of Integer Sequences}
with $0$'s at even indices.  

\subsection{Variable size $1$}
The inserted de Bruijn index has size $1$.  We have
$m_i\, l^1_{n-2,\mathbf{m}}$ where we had
$m_i\, l^{`n}_{n-i-1,\mathbf{m}}$ for natural size.
\begin{eqnarray*}
  l^1_{n+1,0:\mathbf{m}} &=&\sum_{\mathbf{q} `(+)
    \mathbf{r}\,=\,0:\mathbf{m}}\
  \sum_{k=0}^n\, l^1_{k,\mathbf{q}}\,  l^1_{n-k,\mathbf{r}} + 
  \sum_{i=0}^{n-1} (m_i + 1)\ l^1_{n-1,\inc{i}{\mathbf{m}}}\\
  l^1_{n+1,(h+1):\mathbf{m}} &=& \sum_{\mathbf{q} `(+)
    \mathbf{r}=(h+1):\mathbf{m}}\
  \sum_{k=0}^{n}\  l^1_{k,\mathbf{q}}\,  l^1_{n-k,\mathbf{r}}
\end{eqnarray*}

As noticed by Grygiel et al.~\cite{DBLP:journals/logcom/GrygielIZ13}
(§~6.1) There are no linear closed $`l$-terms of size $3k$ and $3k+1$.
However for the values $3k+2$ we get the sequence:
\begin{math}
  1, 5 ,60, 1105, 27120, ...
\end{math}
which is again sequence \textsf{A062980} of the \emph{On-line Encyclopedia
  of Integer Sequences}.  

\section{Counting affine closed terms}
\label{sec:aff}
We have just to add the case \textit{$n\neq 0$ and abstraction without
  binding}. Since no index is added, the size increases by $1$.  The
numbers are written $a^{`n}_{n,\m}$, $a^0_{n,\m}$, $a^1_{n,\m}$, and
$a_{n,\m}$ when the size does not matter.  There are $(0:\m)$-SwissCheeses
of size $n$ that are abstraction without binding.  We get the recursive
formulas:

\subsection{Natural size}
  \begin{eqnarray*}
    a^{`n}_{n+1,0:\mathbf{m}} &=&\sum_{\mathbf{q} `(+)
      \mathbf{r}\,=\,0:\mathbf{m}}\
    \sum_{k=0}^n\, a^{`n}_{k,\mathbf{q}}\,  a^{`n}_{n-k,\mathbf{r}}  + 
    \sum_{i=0}^{p-1} (m_i + 1)\ a^{`n}_{n-i,\inc{i}{\mathbf{m}}}+ a^{`n}_{n,\m}\\
    a^{`n}_{n+1,(h+1):\mathbf{m}} &=& \sum_{\mathbf{q} `(+)
      \mathbf{r}=(h+1):\mathbf{m}}\
    \sum_{k=0}^{n}\  a^{`n}_{k,\mathbf{q}}\,
    a^{`n}_{n-k,\mathbf{r}} 
  \end{eqnarray*}
The numbers of closed affine size with natural size are given in Figure~\ref{fig:until_100}.
\subsection{Variable size $0$}
  \begin{eqnarray*}
    a^0_{n+1,0:\mathbf{m}} &=&\sum_{\mathbf{q} `(+)
      \mathbf{r}\,=\,0:\mathbf{m}}\
    \sum_{k=0}^n\, a^0_{k,\mathbf{q}}\,  a^0_{n-k,\mathbf{r}}  + 
    \sum_{i=0}^{p-1} (m_i + 1)\ a^0_{n,\inc{i}{\mathbf{m}}}+ a^0_{n,\m}\\
    a^0_{n+1,(h+1):\mathbf{m}} &=& \sum_{\mathbf{q} `(+)
      \mathbf{r}=(h+1):\mathbf{m}}\
    \sum_{k=0}^{n}\  a^0_{k,\mathbf{q}}\,
    a^0_{n-k,\mathbf{r}}
  \end{eqnarray*}
The sequence $a^0_{n,0^{`w}}$ of the numbers of affine closed terms for
variable size $0$ is
\begin{displaymath}
  0,1,2,8,29,140,661,3622,19993,120909,744890,4887401,32795272,...
\end{displaymath}
It does not appear in the \emph{On-line Encyclopedia of Integer
  Sequences}.  It corresponds to the coefficients of the  generating function
$\A(z,0)$ where 
\begin{displaymath}
  \A(z,u) = u + z (\A(z,u))^2 + z \frac{\partial \A(z,u)}{\partial u} + z \A(z,u).
\end{displaymath}
\subsection{Variable size $1$}
  \begin{eqnarray*}
    a^1_{n+1,0:\mathbf{m}} &=&\sum_{\mathbf{q} `(+)
      \mathbf{r}\,=\,0:\mathbf{m}}\
    \sum_{k=0}^{n-1}\, a^1_{k,\mathbf{q}}\,  a^1_{n-k,\mathbf{r}}  + 
    \sum_{i=0}^{n-1} (m_i + 1)\ a^1_{n-1,\inc{i}{\mathbf{m}}}+ a^1_{n,\m}\\
    a^1_{n+1,(h+1):\mathbf{m}} &=& \sum_{\mathbf{q} `(+)
      \mathbf{r}=(h+1):\mathbf{m}}\
    \sum_{k=0}^{n}\  a^1_{k,\mathbf{q}}\,
    a^1_{n-k,\mathbf{r}}
  \end{eqnarray*}

The sequence $a^1_{n,0^{`w}}$ of the numbers of affine closed terms for
variable size $1$ is
\begin{displaymath}
0,0,1,2,3,9,30,81,242,838,2799,9365,33616,122937,449698,1696724,6558855, ...
\end{displaymath}
This is sequence \textsf{A281270} in the \emph{On-line Encyclopedia of Integer
  Sequences}.  However it corresponds to the coefficient of the generating
function $\hat{\A}(z,0)$ where $\hat{\A}(z,u)$ is the solution of the
functional equation:
\begin{displaymath}
  \hat{\A}(z,u) = z u + z (\hat{\A}(z,u))^2 + z \frac{\partial \hat{\A}(z,u)}{\partial u} + z \hat{\A}(z,u).
\end{displaymath}
Notice that this corrects the wrong assumptions
of~\cite{DBLP:journals/logcom/GrygielIZ13} (Section~6.2).

\section{Generating functions}
\label{sec:genF}
Consider families $F_{\m}(z)$ of generating functions indexed by $\m$,
where $\m$ is an infinite tuple of naturals.  In fact, we are interested
in the infinite tuples $\m$ that are always $0$, except a finite number of
indices, in order to compute $F_{0^{`w}}(z)$, which corresponds to closed
$`l$-terms.  Let $\uu$ stands for the infinite sequences of variables
$(u_0,u_1,...)$ and $\uu^{\m}$ stands for $(u_0^{m_0}, u_1^{m_1},...,
u_n^{m_n}, ...)$ and $\tail(\uu)$ stand for $(u_1,...)$.  We consider the
series of two variables $z$ and $\uu$ or double series associated with
$F_{\m}(z)$:
\begin{displaymath}
  \mathcal{F}(z,\uu) = \sum_{\m`:\nat^{`w}} F_{\m}(z)\,\uu^{\m}.
\end{displaymath}

\subsection*{Natural size}

$L^{`n}_{\m}(z)$ is associated with the numbers of \emph{closed linear
  SwissCheeses} for natural size:
\begin{eqnarray*}
  L^{`n}_{0:\m} (z) &=&  z\sum_{\m' `(+) \m'' = 0:\m}  L^{`n}_{\m'}(z) L^{`n}_{\m''}(z) + 
  z\, \sum_{i=0}^\infty (m_i+1) z^i L^{`n}_{\inc{i}{\m}}(z)\\
  L^{`n}_{(h+1):\m}(z) &=& [h=0 + \bigwedge_{i=0}^\infty m_i=0] +%
  z \sum_{\m' `(+) \m'' = (h+1):\m} L^{`n}_{\m'}(z) L^{`n}_{\m''}(z)
\end{eqnarray*}
$L^{`n}_{0^{`w}}$ is the generating function for the closed linear
$`l$-terms.  $\L^{`n}(z,\uu)$ is the double series associated with
$L^{`n}_{\m} (z)$ and is solution of the equation:
\begin{eqnarray*}
  \L^{`n}(z,\uu) &=& u_0 + z (\L^{`n}(z,\uu))^2 + %
    \sum_{i=1}^\infty z^{i} \frac{\partial \L^{`n}(z,\tail(\uu))}{\partial u^i}
\end{eqnarray*}
$\L^{`n}(z,0^{`w})$ is the generating function of closed linear
$`l$-terms.

For \emph{closed affine SwissCheeses} we get:
\begin{eqnarray*}
  A^{`n}_{0:\m} (z) &=&  z\sum_{\m' `(+) \m'' = 0:\m}  A^{`n}_{\m'}(z) A^{`n}_{\m''}(z)%
  + z\, \sum_{i=0}^\infty (m_i+1) z^{i} A^{`n}_{\inc{i}{\m}}(z) + %
  z\, A^{`n}_\m(z)\\
  A^{`n}_{(h+1):\m}(z) &=& [h=0 + \bigwedge_{i=0}^\infty m_i=0] + z \sum_{\m' `(+) \m'' = (h+1):\m} A^{`n}_{\m'}(z) A^{`n}_{\m''}(z)
\end{eqnarray*}
$A^{`n}_{0^{`w}}$ is the generating function for the closed linear
$`l$-terms.  $\A^{`n}(z,\uu)$ is the double series associated with
$A^{`n}_{\m} (z)$ and is solution of the equation:
\begin{eqnarray*}
  \A^{`n}(z,\uu) &=& u_0 + z (\A^{`n}(z,\uu))^2 + %
  \sum_{i=1}^\infty z^{i} \frac{\partial
    \A^{`n}(z,\tail(\uu))}{\partial u^i} + z \A^{`n}(z,\tail(\uu))
\end{eqnarray*}
$\A^{`n}(z,0^{`w})$ is the generating function of closed affine
$`l$-terms.

\subsection*{Variable size $0$}

$L^0_{\m}$ is associated with the numbers of \emph{closed linear
  SwissCheeses} for variable size~$0$:
\begin{eqnarray*}
  L^0_{0:\m} (z) &=&  z\sum_{\m' `(+) \m'' = \m}  L^0_{\m'}(z) L^0_{\m''}(z) + 
  z\, \sum_{i=0}^\infty (m_i+1) L^0_{\inc{i}{\m}}(z)\\
  L^0_{(h+1):\m}(z) &=& [h=0 + \bigwedge_{i=0}^\infty m_i=0] + \sum_{\m' `(+) \m'' =
    \m} z L^0_{\m'}(z) L^0_{\m''}(z)
\end{eqnarray*}
$L^0_{0^{`w}}$ is the generating function for the closed linear
$`l$-terms.  $\L^0(z,\uu)$ is the double series associated with $L^0_{\m}
(z)$ and is solution of the equation:
\begin{eqnarray*}
  \L^0(z,\uu) &=& u_0 + z (\L^0(z,\uu))^2 + %
  \sum_{i=1}^\infty \frac{\partial \L^0(z,\tail(\uu)}{\partial u^i}
\end{eqnarray*}
$\L^0(z,0^{`w})$ is the generating function of closed linear $`l$-terms.

For \emph{closed affine SwissCheeses} we get:
\begin{eqnarray*}
  A^0_{0:\m} (z) &=&  z\sum_{\m' `(+) \m'' = 0:\m}  A^0_{\m'}(z) A^0_{\m''}(z)%
  + z\, \sum_{i=0}^\infty (m_i+1) A^0_{\inc{i}{\m}}(z) + %
  z\, A^0_\m(z)\\
  A^0_{(h+1):\m}(z) &=& [h=0 + \bigwedge_{i=0}^\infty m_i=0] + \sum_{\m' `(+) \m'' = (h+1):\m} z A^0_{\m'}(z) A^0_{\m''}(z)
\end{eqnarray*}
$A^0_{0^{`w}}$ is the generating function for the affine linear
$`l$-terms.  $\A^0(z,\uu)$ is the double series associated with $A^0_{\m}
(z)$ and is solution of the equation:
\begin{eqnarray*}
  \A^0(z,\uu) &=& u_0 + z (\A^0(z,\uu))^2 + %
    \sum_{i=1}^\infty \frac{\partial
    \A^0(z,\tail(\uu))}{\partial u^i} + z \A^0(z,\tail(\uu))
\end{eqnarray*}
$\A^0(z,0^{`w})$ is the generating function of closed linear $`l$-terms.
\subsection*{Variable size $1$}
\label{sec:size1}
The generating functions for $l^1_{n,\m}$ are:
\begin{eqnarray*}
  L^1_{0:\m} (z) &=&  z\sum_{\m' `(+) \m'' = \m}  L^1_{\m'}(z) L^1_{\m''}(z) + 
  z^2\, \sum_{i=0}^\infty (m_i+1) L^1_{\inc{i}{\m}}(z)\\
  L^1_{(h+1):\m}(z) &=& [h=0 + \bigwedge_{i=0}^\infty m_i=0] +
                        \sum_{\m' `(+) \m'' = \m} z L^1_{\m'}(z) L^1_{\m''}(z)
\end{eqnarray*}
Then we get as associated double series :
\begin{eqnarray*}
  \L^1(z,\uu) &=& u_0 + z (\L^1(z,\uu))^2 + %
  z^2  \sum_{i=1}^\infty \frac{\partial \L^1(z,\tail(\uu))}{\partial u^i}
\end{eqnarray*}

\section{Effective computations}
\label{sec:comp}

The definition of the coefficients $a^{`n}_{\m}$ and others is highly
recursive and requires a mechanism of memoization. In Haskell, this can be
done by using the call by need which is at the core of this language.
Assume we want to compute the values of $a^{`n}_{\m}$ until a value
\texttt{upBound} for $n$.  We use a recursive data structure:
\begin{minted}{haskell}
data Mem = Mem [Mem] | Load [Integer]
\end{minted}
in which we store the computed values of a function
\begin{minted}{haskell}
a :: Int -> [Int] -> Integer
\end{minted}
In our implementation the depth of the recursion of \pourpre{\texttt{Mem}}
is limited by \texttt{upBound}, which is also the longest tuple $\m$ for
which we will compute $a^{`n}_{\m}$.  Associated with
\pourpre{\texttt{Mem}} there is a function
\begin{minted}{haskell}
access :: Mem -> Int -> [Int] -> Integer access (Load l) n [] = l !! n
access (Mem listM) n (k:m) = access (listM !! k) n m
\end{minted}
The leaves of the tree memory, corresponding to \pourpre{\texttt{Load}},
contains the values of the function:
\begin{minted}{haskell}
memory :: Int -> [Int] -> Mem 
memory 0 m = Load [a n (reverse m) | n<-[0..]]  
memory k m = Mem [memory (k-1) (j:m) | j<-[0..]]
\end{minted}
The memory relative to the problem we are interested in is
\begin{minted}{haskell}
theMemory = memory (bound) []
\end{minted}
and the access to \bl{\texttt{theMemory}} is given by a specific function:
\begin{minted}{haskell}
acc :: Int -> [Int] -> Integer acc n m = access theMemory n m
\end{minted}
Notice that \bl{\texttt{a}} and \bl{\texttt{acc}} have the same
signature.  This is not a coincidence, since \bl{\texttt{acc}} accesses
values of \bl{\texttt{a}} already computed.  Now we are ready to express
\bl{\texttt{a}}:
\begin{minted}{haskell}
a 0 m = iv (head m == 1 && all ((==) 0) (tail m)) 
a n m = aAPP n m + aABSwB n m + aABSnB n m
\end{minted}
\bl{\texttt{aAPP}} \emph{counts affine terms that are applications:}
\ifHAL
\begin{minted}{haskell}
aAPP n m = sum (map (\textbackslash((q,r),(k,nk))->(acc k q)*(acc nk r)) (allCombinations m (n-1)))
\end{minted}
\else
\begin{minted}{haskell}
aAPP n m = sum (map (\((q,r),(k,nk))->(acc k q)*(acc nk r))
                     (allCombinations m (n-1)))
\end{minted}
\fi where \bl{\texttt{allCombinations}} returns a list of all the pairs of
pairs $(\m',\m'')$ such $\m = \m' `(+) \m''$ and of pairs $(k,nk)$ such
that $k+nk = n$.  \bl{\texttt{aABSwB}} \emph{counts affine terms that are
  abstractions with binding}.
\begin{minted}{haskell}
aABSwB n m 
   | head m == 0 = sum [aABSAtD n m i |i<-[1..(n-1)]] 
   | otherwise = 0
\end{minted}
\bl{\texttt{aABSAtD}} \emph{counts affine terms that are abstractions
  with binding at level $i$}:
\begin{minted}{haskell}
aABSAtD n m i = (fromIntegral (1 + m!!i))*(acc (n-i-1) (tail (inc i m) ++ [0]))
\end{minted}
\bl{\texttt{aABSnB}} \emph{counts affine terms that are abstractions with
  no binding}:
\begin{minted}{haskell}
aABSnB n m 
     | head m == 0 = (acc (n-1) (tail m ++ [0])) 
     | otherwise = 0
\end{minted}
Anyway the efficiency of this program is limited by the size of the
memory, since for computing $a^{`n}_{n,0^{n}}$, for instance, we need to
compute $a^{`n}_{\mathbf{r}}$ for about $n!$ values.

\section{Generating affine and linear terms}
\label{sec:gen}

By relatively small changes it is possible to build programs which
generate linear and affine terms.  For instance for generating affine
terms we get.

\ifHAL
\begin{minted}{haskell}
amg :: Int -> [Int] -> [SwissCheese]
amg 0 m =  if (head m == 1 && all ((==) 0) (tail m)) then [Box 0] else []
amg n m = allAPP n m ++ allABSwB n m ++ allABSnB n m 

allAPP  :: Int -> [Int] -> [SwissCheese]
allAPP n m = foldr (++) [] (map (\textbackslash((q,r),(k,nk))-> appSC (cartesian (accAG k q)
                                                                   (accAG nk r))
                            (allCombinations m (n-1)))

allABSAtD :: Int -> [Int] -> Int -> [SwissCheese]
allABSAtD n m i = foldr (++) [] (map (abstract (i-1)) (accAG (n - i - 1)
                                                      (tail (inc i m) ++ [0])))
                  
allABSwB  :: Int -> [Int] -> [SwissCheese]
allABSwB n m 
  | head m == 0 = foldr (++) [] [allABSAtD n m i |i<-[1..(n-1)]]
  | otherwise = []

allABSnB  :: Int -> [Int] -> [SwissCheese]
allABSnB n m
  | head m == 0 = map (AbsSC . raise) (accAG (n-1) (tail m ++ [0]))
  | otherwise = []

memoryAG :: Int -> [Int] -> MemSC
memoryAG 0 m = LoadSC [amg n (reverse m) | n<-[0..]]
memoryAG k m = MemSC [memoryAG (k-1) (j:m) | j<-[0..]] 

theMemoryAG = memoryAG (upBound) []

accAG :: Int -> [Int] -> [SwissCheese]
accAG n m = accessSC theMemoryAG n m
\end{minted}
\else
\begin{minted}{haskell}
amg :: Int -> [Int] -> [SwissCheese]
amg 0 m =  if (head m == 1 && all ((==) 0) (tail m)) then [Box 0] else []
amg n m = allAPP n m ++ allABSwB n m ++ allABSnB n m 

allAPP  :: Int -> [Int] -> [SwissCheese]
allAPP n m = foldr (++) [] (map (\((q,r),(k,nk))-> appSC (cartesian (accAG k q)
                                                                    (accAG nk r))
                            (allCombinations m (n-1)))

allABSAtD :: Int -> [Int] -> Int -> [SwissCheese]
allABSAtD n m i = foldr (++) [] (map (abstract (i-1)) (accAG (n - i - 1)
                                                      (tail (inc i m) ++ [0])))
                  
allABSwB  :: Int -> [Int] -> [SwissCheese]
allABSwB n m 
  | head m == 0 = foldr (++) [] [allABSAtD n m i |i<-[1..(n-1)]]
  | otherwise = []

allABSnB  :: Int -> [Int] -> [SwissCheese]
allABSnB n m
  | head m == 0 = map (AbsSC . raise) (accAG (n-1) (tail m ++ [0]))
  | otherwise = []

memoryAG :: Int -> [Int] -> MemSC
memoryAG 0 m = LoadSC [amg n (reverse m) | n<-[0..]]
memoryAG k m = MemSC [memoryAG (k-1) (j:m) | j<-[0..]] 

theMemoryAG = memoryAG (upBound) []

accAG :: Int -> [Int] -> [SwissCheese]
accAG n m = accessSC theMemoryAG n m
\end{minted}
\fi %
From this, we get programs for generating random affine terms or random
linear terms as follows: if we want a random closed linear term of a given
size $n$, we throw a random number, say $p$, between $1$ and $l_{n,0^n}$
and we look for the $p^{th}$ in the list of all the closed linear terms of
size $n$.  Haskell laziness mimics the unranking. Due to high requests in
space, we cannot go further than the random generation of closed linear
terms of size $23$ and closed affine terms of size $19$ . There are similar programs for generating all the
terms of size $n$ for variable size $0$ and variable size $1$.

\section{Normal forms}
\label{sec:nf}

From the method used for counting affine and linear closed terms, it is
easy to deduce method for counting affine and linear closed normal forms.
Like before, we use SwissCheeses.  In this section we consider only
natural size.

\subsection{Natural size}
\label{sec:nfNatural}

\subsubsection*{Affine closed normal forms}
Let us call $anf^{`n}_{n,\m}$ the numbers of affine SwissCheeses with no
$`b$-redex and $ane^{`n}_{n,\m}$ the numbers of neutral affine SwissCheeses,
i.e., affine SwissCheeses with no $`b$-redexes that are sequences of
applications starting with a de Bruijn index.  In addition we count:
\begin{itemize}
\item $anf^{`n}`lw_{n,m}$ the number of affine SwissCheeses with no $`b$-redex
  which are abstraction with a binding of a de Bruijn index,
\item $anf^{`n}`ln_{n,m}$ the number of affine SwissCheeses with no $`b$-redex
  which are abstraction with no binding.
\end{itemize}

\begin{eqnarray*}
  anf^{`n}_{0,\m} &=&  ane^{`n}_{0,\m}\\
  anf^{`n}_{n+1,\m} &=& ane^{`n}_{n+1,\m} + anf^{`n}`lw_{n+1,m} + anf^{`n}`ln_{n+1,m}
\end{eqnarray*}
where
\begin{eqnarray*}
  ane^{`n}_{0,\m} &=& m_0 = 1 \wedge \bigwedge_{j=1}^{p-1} m_j = 0\\
  ane^{`n}_{n+1,\m} &=& \sum_{\mathbf{q} `(+)
    \mathbf{r}\,=\,0:\mathbf{m}}\
  \sum_{k=0}^n\, ane^{`n}_{k,\mathbf{q}}\,  anf^{`n}_{n-k,\mathbf{r}}
\end{eqnarray*}
and
\begin{eqnarray*}
  anf^{`n}`lw_{n+1,m} &=&  \sum_{i=0}^{n} (m_i + 1)\ anf^{`n}_{n-i,\inc{i}{\mathbf{m}}}
\end{eqnarray*}
and
\begin{eqnarray*}
  anf^{`n}`ln_{n+1,m} &=& anf^{`n}_{n,m}
\end{eqnarray*}
There are two generating functions, $\ANF$ and $\ANE$, which are
associated to $anf^{`n}_{n,\m}$ and $anf^{`n}_{n,\m}$:
\begin{eqnarray*}
  \ANF(z,\uu)&=&  \ANE(z,\uu) + %
    \sum_{i=1}^\infty z^{i} \frac{\partial
    \ANF(z,\tail(\uu))}{\partial u^i} + z \ANF(z,\tail(\uu))\\
  \ANE(z,\uu)&=&  u_0 + z \ANE(z,\uu) \ANF(z,\uu) 
\end{eqnarray*}
\subsubsection*{Linear closed normal forms}
Let us call $lnf^{`n}_{n,\m}$ the numbers of linear SwissCheeses with no
$`b$-redex and $lne^{`n}_{n,\m}$ the numbers of neutral linear SwissCheeses,
linear SwissCheeses with no $`b$-redexes that are sequences of
applications starting with a de Bruijn index.  In addition we count
$lnf^{`n}`lw_{n,m}$ the number of linear SwissCheeses with no $`b$-redex which
are abstraction with a binding of a de Bruijn index.

\begin{eqnarray*}
  lnf^{`n}_{0,\m} &=&  lne^{`n}_{0,\m}\\
  lnf^{`n}_{n+1,\m} &=& lne^{`n}_{n+1,\m} + lnf^{`n}`lw_{n+1,m}
\end{eqnarray*}
where
\begin{eqnarray*}
  lne^{`n}_{0,\m} &=& m_0 = 1 \wedge \bigwedge_{j=1}^{p-1} m_j = 0\\
  lne^{`n}_{n+1,\m} &=& \sum_{\mathbf{q} `(+)
    \mathbf{r}\,=\,0:\mathbf{m}}\
  \sum_{k=0}^n\, lne^{`n}_{k,\mathbf{q}}\,  lnf^{`n}_{n-k,\mathbf{r}}
\end{eqnarray*}
and
\begin{eqnarray*}
  lnf^{`n}`lw_{n+1,m} &=&  \sum_{i=0}^{p-1} (m_i + 1)\ lnf^{`n}_{n-i,\inc{i}{\mathbf{m}}}
\end{eqnarray*}
with the two generating functions:
\begin{eqnarray*}
  \LNF(z,\uu)&=&  \LNE(z,\uu) + %
   \sum_{i=1}^\infty z^{i} \frac{\partial
    \LNF(z,\tail(\uu))}{\partial u^i} \\
  \LNE(z,\uu)&=&  u_0 + z \LNE(z,\uu) \LNF(z,\uu) 
\end{eqnarray*}
We also deduce programs for generating all the closed affine or linear
normal forms of a given size from which we deduce programs for \emph{generating
random closed affine or linear normal forms of a given size}.  For
instance, here are three randoms linear closed normal forms (using de
Bruijn indices) of natural size~$28$:
\begin{small}
  \begin{displaymath}
    `l `l `l `l (2~ `l ((1~2)~ `l (0~ (5~1))))
    \quad
    `l (0~ `l `l (1~`l `l ((0~ (2~`l `l ((1~`l 0)~0)))~ 1)))
    \quad
    `l ((0~`l 0)~ `l `l ((0 ((1~`l 0) `l `l (1~(0~`l 0)))) `l 0))  
  \end{displaymath}
\end{small}

\subsection{Variable size $0$}
\label{sec:nf0}

\subsubsection*{Linear closed normal forms}

A little like previously, let us call $lnf^0_{n,\m}$ the numbers of linear SwissCheeses with no
$`b$-redex and $lne^0_{n,\m}$ the numbers of neutral linear SwissCheeses,
linear SwissCheeses with no $`b$-redexes that are sequences of
applications starting with a de Bruijn index.  In addition we count
$lnf^0`lw_{n,m}$ the number of linear SwissCheeses with no $`b$-redex which
are abstraction with a binding of a de Bruijn index. We assume that the
reader knows now how to proceed.

\begin{eqnarray*}
  lnf^0_{0,\m} &=& lne^0_{0,\m}\\
  lnf^0_{n+1,\m} &=& lne^0_{n+1,\m} + lnf^0`lw_{n+1,m}
\end{eqnarray*}
where
\begin{eqnarray*}
  lne^0_{0,\m} &=& m_0 = 1 \wedge \bigwedge_{j=1}^{p-1} m_j = 0\\
  lne^0_{n+1,\m} &=& \sum_{\mathbf{q} `(+)
    \mathbf{r}\,=\,0:\mathbf{m}}\
  \sum_{k=0}^n\, lne^0_{k,\mathbf{q}}\,  lnf^0_{n-k,\mathbf{r}}
\end{eqnarray*}
\begin{eqnarray*}
  lnf^0`lw_{n+1,m} &=&  \sum_{i=0}^n (m_i + 1)\ lnf^0_{n,\inc{i}{\mathbf{m}}}
\end{eqnarray*}
 and the
two generating functions:
\begin{eqnarray*}
  \LNFZ(z,\uu)&=&  \LNEZ(z,\uu) + %
   \sum_{i=1}^\infty \frac{\partial
    \LNFZ(z,\tail(\uu))}{\partial u^i} \\
  \LNEZ(z,\uu)&=&  u_0 + z \LNEZ(z,\uu) \LNFZ(z,\uu) 
\end{eqnarray*}
With no surprise we get for $lnf^0_{n,0^{n}}$ the sequence:
\begin{displaymath}
  0,1,0,3,0,26,0,367,0,7142,0,176766,0,5304356, ...
\end{displaymath}
mentioned by Zeilberger in~\cite{DBLP:journals/corr/Zeilberger15} and listing
the coefficients of the generating function $\LNFZ(z,0^{`w})$.

We let the reader deduce how to count closed affine normal forms  for variable size $0$ and
closed linear and affine normal forms for variable size $1$ alike.  Notice
that the Haskell programs are on the GitHub site.

\section{Related works and Acknowledgement}
There are several works on counting $`l$-terms, for instance on natural
size~\cite{DBLP:conf/sofsem/BendkowskiGLZ16,DBLP:journals/corr/BendkowskiGLZ16},
on variable
size~$1$~\cite{gittenberger-2011-ltbuh,DBLP:journals/lmcs-2009,DBLP:journals/tcs/Lescanne13},
on variable size $0$~\cite{DBLP:journals/jfp/GrygielL13}, on affine terms
with variable size~1~\cite{DBLP:journals/tcs/BodiniGJ13,DBLP:journals/combinatorics/BodiniGGJ13}, on linear
$`l$-terms~\cite{DBLP:journals/corr/ZeilbergerG14,DBLP:journals/corr/Zeilberger15,DBLP:journals/jfp/Zeilberger16}, also on a size based
binary representation of the
$`l$-calculus~\cite{DBLP:journals/jfp/GrygielL15}
(see~\cite{DBLP:conf/stacs/GittenbergerG16} for a synthetic view of both
natural size and binary size).

The basic idea of this work comes from a discussion with Maciej
Bendkowski, Olivier Bodini, Sergey Dovgal and Katarzyna Grygiel, I thank
them as I thank Noam Zeilberger for interactions.

\section{Conclusion}
\label{sec:conc}

This presentation shares similarity with this of
~\cite{DBLP:journals/jfp/GrygielL13,DBLP:journals/jfp/GrygielL15,DBLP:conf/padl/BendkowskiGT17}. Instead
of considering the size $n$ and the bound $m$ of free indices like in
expressions of the form:
\begin{displaymath}
  T_{n+1,m} = T_{n,m+1} + \sum_{i=0}^n T_{i,m} T_{n-i,m}
\end{displaymath}
here we replace $m$ by the characteristic $\m$.  We can imagine a common
framework. On another hand, as noticed by Paul Tarau, this approach has
features of dynamic programming~\cite{Cormen:2009:IAT:1614191}, which
makes it somewhat efficient.


\appendix
\section*{Data}
\label{sec:data}
In the appendix, we give the first values of
$l^{`n}_{n,0^n}$, $a^{`n}_{n,0^n}$, and $anf^{`n}_{n,0^n}$.
\begin{figure}[t!]
  \centering
  \begin{footnotesize}
    \begin{displaymath}
      \begin{array}[h]{ll}
        0 & 0\\
        1 & 0\\
        2 & 1\\
        3 & 0\\
        4 & 0\\
        5 & 3\\
        6 & 2\\
        7 & 0\\
        8 & 16\\
        9 & 24\\
        10 & 8\\
        11 & 117\\
        12 & 252\\
        13 & 180\\
        14 & 1024\\
        15 & 2680\\
        16 & 2952\\
        17 & 10350\\
        18 & 29420\\
        19 & 42776\\
        20 & 116768\\
        21 & 335520\\
        22 & 587424\\
        23 & 1420053\\
        24 & 3976424\\
        25 & 7880376\\
        26 & 18103936\\
        27 & 48816576\\
        28 & 104890704\\
        29 & 237500826\\
        30 & 617733708\\
        31 & 1396750576\\
        32 & 3171222464\\
        33 & 8014199360\\
        34 & 18688490336\\
        35 & 42840683418\\
        36 & 106063081288\\
        37 & 251769197688\\
        38 & 583690110208\\
        39 & 1425834260080\\
        40 & 3417671496432\\
        41 & 8007221710652\\
        42 & 19404994897976\\
        43 & 46747189542384\\
        44 & 110498345360800\\
        45 & 266679286291872\\
        46 & 644021392071840\\
        47 & 1533054190557133\\
        48 & 3693823999533360\\
        49 & 8931109667692464 \\
        50 & 21375091547312128
      \end{array}
      \qquad
      \begin{array}[h]{ll}
        51 & 51496022711337536\\
        52 & 124591137939086496\\
        53 & 299402908258405410\\
        54 & 721839933329222924\\
        55 & 1747307145272084192\\
        56 & 4211741383777966592\\
        57 & 10165998012602469888\\
        58 & 24620618729658655936\\
        59 & 59482734150603634286\\
        60 & 143764591607556354344\\
        61 & 348379929166234350008\\
        62 & 843169238563254723200\\
        63 & 2040572920613086128400\\
        64 & 4948102905207104837424\\
        65 & 11992521016286173712196\\
        66 & 29059897435554891991144\\
        67 & 70516464312280927105392\\
        68 & 171105110698292441423968\\
        69 & 415095704639682396539232\\
        70 & 1008016383720573882885792\\
        71 & 2448305474519849567597826\\
        72 & 5945721872300885649415632\\
        73 & 14449388516068567845838736\\
        74 & 35125352062243788817753856\\
        75 & 85382289240293493116120064\\
        76 & 207650379931166057815603296\\
        77 & 505172267243918348155299780\\
        78 & 1229005880128485245247395000\\
        79 & 2991079243470267667831893408\\
        80 & 7281852742753184123608419712\\
        81 & 17729171587798767750815341440\\
        82 & 43177454620325445122944305984\\
        83 & 105185452787117035266315446868\\
        84 & 256273862465425158211948020048\\
        85 & 624527413292252904584121980208\\
        86 & 1522355057007327280427270436480\\
        87 & 3711429775030704772089070886624\\
        88 & 9050041253711022076275958636128\\
        89 & 22073150301758857110072042919800\\
        90 & 53844910909398928990641101351664\\
        91 & 131371135544173914537076774932576\\
        92 & 320588677238085642820920910555968\\
        93 & 782465218885869813183863213231424\\
        94 & 1910077425906069707804966102543936\\
        95 & 4663586586924802791117231052636349\\
        96 & 11388259565942452837717688743953504\\
        97 & 27813754361897984543467478917223008\\
        98 & 67941781284113201998645699501746176\\
        99 & 165989485724048964272023600773271424\\
        100 & 405588809305168453963137377442321728
      \end{array}
    \end{displaymath}
  \end{footnotesize}
  \caption{\emph{Natural size:} numbers of closed linear terms of size $n$ from
    $0$ to $100$}
  \label{fig:lin_until_10}
\end{figure}
\begin{figure}[th!]
  \centering
  \begin{footnotesize}
    \begin{displaymath}
      \begin{array}[h]{ll}
        0 & 0\\
        1 & 0\\
        2 & 1\\
        3 & 1\\
        4 & 2\\
        5 & 5\\
        6 & 12\\
        7 & 25\\
        8 & 64\\
        9 & 166\\
        10 & 405\\
        11 & 1050\\
        12 & 2763\\
        13 & 7239\\
        14 & 19190\\
        15 & 51457\\
        16 & 138538\\
        17 & 374972\\
        18 & 1020943\\
        19 & 2792183\\
        20 & 7666358\\
        21 & 21126905\\
        22 & 58422650\\
        23 & 162052566\\
        24 & 450742451\\
        25 & 1256974690\\
        26 & 3513731861\\
        27 & 9843728012\\
        28 & 27633400879\\
        29 & 77721141911\\
        30 & 218984204904\\
        31 & 618021576627\\
        32 & 1746906189740\\
        33 & 4945026080426\\
        34 & 14017220713131\\
        35 & 39784695610433\\
        36 & 113057573020242\\
        37 & 321649935953313\\
        38 & 916096006168770\\
        39 & 2611847503880831\\
        40 & 7453859187221508\\
        41 & 21292177500898858\\
        42 & 60875851617670699\\
        43 & 174195916730975850\\
        44 & 498863759031591507\\
        45 & 1429753835635525063\\
        46 & 4100730353324163138\\
        47 & 11769771167532816128\\
        48 & 33804054749367200891\\
        49 & 97151933333668422006\\
        50 & 279385977720772581435\\
      \end{array}
      \qquad
      \begin{array}[h]{ll}
        51 & 803928779462727941247\\
        52 & 2314623127904669382002\\
        53 & 6667810436356967142481\\
        54 & 19218411059885449257096\\
        55 & 55421020161661024650870\\
        56 & 159899218321197381984561\\
        57 & 461557020400062903560120\\
        58 & 1332920908954281811200519\\
        59 & 3851027068336583693412910\\
        60 & 11131032444503136571789527\\
        61 & 32186581221116996967632029\\
        62 & 93108410048006285466998584\\
        63 & 269446191702411420790402033\\
        64 & 780043726186403167392453886\\
        65 & 2259043189995515315930349650\\
        66 & 6544612955390252336187266873\\
        67 & 18966737218108971681014445025\\
        68 & 54985236298270057405776629352\\
        69 & 159455737350384637847783055311\\
        70 & 462562848624435724964181323484\\
        71 & 1342251884451664733064283251627\\
        72 & 3896065622127200625653134100538\\
        73 & 11312117748805772104795220337816\\
        74 & 32853646116456632492645965741531\\
        75 & 95442534633482460553801961967438\\
        76 & 277342191547330839640289978813667\\
        77 & 806125189457291902863848267463755\\
        78 & 2343682130911232279285707290604156\\
        79 & 6815564023736534208079367816340359\\
        80 & 19824812322145727566417303371819466\\
        81 & 57679033022808238913186144092831856\\
        82 & 167851787082561392384648248846390041\\
        83 & 488574368670832093243802790464796207\\
        84 & 1422426342380883254459783410845365006\\
        85 & 4142104564089044203901190817275864665\\
        86 & 12064305885705003967881526911560653106\\
        87 & 35145647815239737143373764367447378676\\
        88 & 102406303052123097062053564818109468705\\
        89 & 298446029598661205216170897850336550644\\
        90 & 869935452705023302189031644932803990417\\
        91 & 2536229492704354513309696228592784181158\\
        92 & 7395518143425160073537967606298755947391\\
        93 & 21568776408467701927134211542478146593789\\
        94 & 62915493935623036562559989770249004382816\\
        95 & 183553775888862113259168150130266362416356\\
        96 & 535600661621556969155453544692826625532079\\
        97 & 1563109720672526919899689366626240867515144\\
        98 & 4562542818801138452310024131223304186909233\\
        99 & 13319630286623965617386598746472280781972745\\
        100& 38890520391341859449843201188612375394153776
      \end{array}
    \end{displaymath}
  \end{footnotesize}
  \caption{\emph{Natural size:} numbers of closed affine terms of size $n$ from
    $0$ to $100$}
  \label{fig:until_100}
\end{figure}
\begin{figure}[!tbp]
  \centering
  \begin{footnotesize}
    \begin{displaymath}
      \begin{array}[c]{ll}
        0 & 0\\
        1 & 0\\
        2 & 1\\
        3 & 1\\
        4 & 2\\
        5 & 3\\
        6 & 7\\
        7 & 10\\
        8 & 20\\
        9 & 40\\
        10 & 77\\
        11 & 160\\
        12 & 318\\
        13 & 671\\
        14 & 1405\\
        15 & 2981\\
        16 & 6312\\
        17 & 13672\\
        18 & 29399\\
        19 & 63697\\
        20 & 139104\\
        21 & 304153\\
        22 & 667219\\
        23 & 1469241\\
        24 & 3247176\\
        25 & 7184288\\
        26 & 15949179\\
        27 & 35480426\\
        28 & 79083472\\
        29 & 176607519\\
        30 & 395119875\\
        31 & 885450388\\
        32 & 1987289740\\
        33 & 4466760570\\
        34 & 10053371987\\
        35 & 22656801617\\
        36 & 51121124910\\
        37 & 115478296639\\
        38 & 261139629999\\
        39 & 591138386440\\
        40 & 1339447594768
      \end{array}
      \qquad
      \begin{array}[h]{ll}
        41 & 3037843646560\\
        42 & 6895841598615\\
        43 & 15666498585568\\
        44 & 35620848278448\\
        45 & 81052838239593\\
        46 & 184564847153821\\
        47 & 420564871255118\\
        48 & 958975854646984\\
        49 & 2188068392529104\\
        50 & 4995528560788451\\
        51 & 11411921511827547\\
        52 & 26084524952754538\\
        53 & 59654682828889245\\
        54 & 136500653558490261\\
        55 & 312496493161999851\\
        56 & 715760763686417314\\
        57 & 1640194881084692664\\
        58 & 3760284787917366081\\
        59 & 8624561382605096780\\
        60 & 19789639944299656346\\
        61 & 45427337308377290201\\
        62 & 104320438668034814453\\
        63 & 239656248361374562433\\
        64 & 550769764273325683828\\
        65 & 1266217774600330829940\\
        66 & 2912050679107531357883\\
        67 & 6699418399886008666265\\
        68 & 15417663698156810292010\\
        69 & 35492710197462925262295\\
        70 & 81732521943462960197057\\
        71 & 188270363628099910161436\\
        72 & 433807135012774797924026\\
        73 & 999851681931974600766994\\
        74 & 2305129188866501774481545\\
        75 & 5315847675735178072941600\\
        76 & 12262083079763320881047944\\
        77 & 28292248892584567512609357\\
        78 & 65294907440089718078048829\\
        79 & 150729070403767032817820543\\
        80 & 348031015577337732605480908
      \end{array}
    \end{displaymath}
  \end{footnotesize}
  \caption{\emph{Natural size:} numbers of closed affine normal forms of size
    $n$ from $0$ to $80$}
  \label{fig:acnf}
\end{figure}
\end{document}